\documentclass[epj]{svjour}
\usepackage{graphics}
\usepackage{amsmath}
\usepackage{amssymb}
\begin{document}
\title{R-mode astronomy}
\author{Kostas.~D.~Kokkotas \and
Kai Schwenzer
}                     
%
%
\institute{Theoretical Astrophysics (IAAT), Eberhard Karls University of T\"ubingen, T\"ubingen 72076, Germany
}
\date{Received: date / Revised version: date}
%
\abstract{
Next generation gravitational wave detectors will start taking data in the near future. Here we discuss the chances to detect the continuous emission from r-mode oscillations in compact stars and study which properties of compact stars we can infer from such novel data. In particular we show that the combination of the gravitational wave data with electromagnetic multi-messenger observations could give us detailed insight into compact star properties, ranging from precise mass-radius 
measurements to the determination of the equation of state and the phase structure of dense matter.
} 
\maketitle
\section{Introduction}
\label{intro}

Gravitational waves from global oscillation modes provide the only direct way to probe the opaque interior of compact stars. The key reason why global modes are a very promising source of gravitational waves is that general relativistic hydrodynamics predicts the instability of a class of modes due to the Friedman-Schutz mechanism \cite{Friedman:1978hf}. I.e. these modes will be present unless the dissipation in the interior is strong enough to prevent this, providing us with valuable information on the star's composition. R-modes \cite{Papaloizou:1978zz,Andersson:1997xt,Friedman:1997uh,Lindblom:1998wf,Andersson:1998ze,Andersson:2000mf} are particularly promising for gravitational wave astroseismology, since in the absence of dissipation they would be unstable at all rotation frequencies. Even in the presence of viscous damping they could be unstable in sources that spin with frequencies down to below 100 Hz, so that a large number of sources at various stages in their evolution could emit continuous gravitational waves. F-modes could also be an interesting source of gravitational waves \cite{Stergioulas:2003yp,Passamonti:2011eh,Doneva:2015jaa}, but since they are not unstable at frequencies far below the Kepler limit of the star \cite{Passamonti:2012yx}, they should only be present in newly-formed sources.

The theoretical study of the gravitational wave emission due to r-modes has been pioneered in \cite{Owen:1998xg} and this topic has received enhanced interest in recent years \cite{Bondarescu:2008qx,Owen:2010ng,Alford:2012yn,Mahmoodifar:2013quw,Alford:2014pxa,Mytidis:2015} due to the advent of strongly improved second generation detectors, like the advanced LIGO \cite{Harry:2010zz,TheLIGOScientific:2014jea}, the advanced Virgo \cite{TheVirgo:2014hva} or the KAGRA \cite{Aso:2013eba} interferometers. Most previous searches for continuous gravitational waves were aimed exclusively at star deformations parametrized by an ellipticity \cite{Abbott:2003yq,Abbott:2008fx,Wette:2008hg,Collaboration:2009rfa,Aasi:2013sia}. However, r-mode oscillations have been identified as viable and promising targets for continuous gravitational wave searches \cite{Owen:2010ng} and have been taken into account in recent analyses \cite{Abadie:2010hv,Aasi:2014ksa}.
Since gravitational waves could be detected in the near future, it is timely to ask what information about compact sources we could gain from a r-mode gravitational wave signal \cite{Weinstein:2011kh}. Such a detection would open the novel field of r-mode gravitational wave astroseismology, or {\em r-mode astronomy} for short, and perfectly extend our present spectral window to the cosmos towards the sub-kHz regime. This novel form of astronomy would allow us to directly probe the interior of dense sources, and in this work we will sketch its basic aspects. 

R-mode gravitational wave emission, discussed in sec. \ref{sec:r-mode-emission}, depends strongly on the structure and evolution of compact stars and we will discuss the scenarios under which r-modes can be present in sec. \ref{sec:scenarios}. Whereas in initial studies it was assumed that r-modes of the maximal amplitude allowed by the structural stability of the star could be present \cite{Owen:1998xg}, in recent years restrictive bounds on the r-mode amplitude were set by electromagnetic observations \cite{Mahmoodifar:2013quw,Alford:2013pma}. It is encouraging that despite these strict limits it turned out that several sources could be above or close to the sensitivity of next generation detectors and many dozens of sources should be in reach with the advent of third generation facilities \cite{Alford:2012yn,Mahmoodifar:2013quw,Alford:2014pxa,Mytidis:2015}, as is surveyed in sec. \ref{sec:detectability}. A continuous gravitational wave detection would require to discriminate r-mode emission from other emission mechanisms and to distinguish the different r-mode emission scenarios. We will discuss the identification and the direct observables that can be obtained in sec. \ref{sec:identification}. Finally, we show in sec. \ref{sec:multi-messenger} that even considerably more information about a sources can be obtained from multi-messenger observations, where the gravitational wave signal is combined with electromagnetic data. 

\section{Gravitational wave emission from r-modes}
\label{sec:r-mode-emission}

\subsection{R-mode oscillations}
Classical r-modes \cite{Papaloizou:1978zz,Andersson:1997xt,Andersson:2000mf} are purely axial modes of a rotating star. They are characterized by a velocity perturbation in non-radial direction given in terms of magnetic type vector spherical harmonics with $m\!=\!l$ and are in the non-relativistic limit given by\footnote{The form of the prefactor of the angular dependence, especially the appearance of the rotation angular velocity $\Omega$ instead of the oscillation angular velocity $\omega$, determines the particular definition of the r-mode amplitude $\alpha$.} 
\begin{equation}
\delta \vec v=\alpha\Omega Rf_l\!\left(\frac{r}{R}\right) \vec Y^{B}_{ll}(\theta,\phi) e^{i\omega t} \, , \label{eq:r-mode}
\end{equation}
where $\alpha$ is the dimensionless r-mode amplitude and $f_l$ is an arbitrary dimensionless radial profile function. Although all these modes are generically unstable, the fundamental $l\!=\!2$ mode couples most strongly to gravitational waves. Therefore, it has by far the smallest growth time \cite{Owen:1998xg}, so that in fast-rotating sources it cannot be damped by standard viscous mechanisms \cite{Lindblom:1998wf,Alford:2010fd}. At the same time it dominates the gravitational wave emission \cite{Owen:1998xg} and is therefore the relevant mode for gravitational wave searches. 

R-modes in realistic stars are in general so-called ``axial-led'' hybrid modes \cite{Lockitch:1998nq}, where the axial component dominates but also polar admixtures are present. Their relativistic generalization has been studied in \cite{Lockitch:2000aa} and is therefore described by more complicated expressions for the appropriate covariant 4-velocity involving additional radial profile functions. However, fortunately in explicit numeric relativistic analyses it turned out that, even though relativistic corrections to static star properties are sizable, the corrections to the mode eigenfunctions are tiny and for the fundamental r-mode the Newtonian expression eq.~(\ref{eq:r-mode}) agrees to about 1\% accuracy with the corresponding relativistic result \cite{Lockitch:2000aa,Lockitch:2002sy,Idrisy:2014qca}. Therefore, we will in following use the simple non-relativistic form for the r-mode eq.~(\ref{eq:r-mode}), noting the small uncertainty this induces on the results presented below. 

Relativistic corrections have a significant impact on the r-mode spectrum, though. In Newtonian approximation and for slowly rotating stars the oscillation angular velocity  in the inertial frame $\omega\!=\!2\pi \nu$ and the rotation angular velocity $\Omega\!=\!2\pi f$ of the fundamental r-mode are canonically related by $\omega\!=\!-4/3 \Omega$ and the radial profile takes the form $f_2(r)\!=\!r^2$. Both relativistic corrections \cite{Lockitch:2000aa} and rotation corrections in fast-rotating sources \cite{Lindblom:1999yk,Gaertig:2008uz} alter this simple form and in general also introduce a dependence on the star configuration determined by the equation of state of dense matter. This is usually described by a multiplicative function $\kappa(\Omega)$ encoding the connection in the rotating frame, but in the following it will be convenient to describe it instead by a correction factor $\chi(\Omega)$ that parametrizes the deviation from the canonical relation in the inertial frame
\begin{equation}
\omega = \left( \kappa(\Omega)-2 \right) \Omega \equiv - \frac{4}{3} \chi (\Omega) \Omega\label{eq:gw-frequency}
\end{equation}
where $\chi (\Omega) \!\approx\! 1$ describes the corrections due to relativity \cite{Lockitch:2000aa,Lockitch:2002sy,Idrisy:2014qca} (increasing it) and rotation \cite{Lindblom:1999yk} (decreasing it).
In general this relation is complicated and its determination requires numerical general relativistic hydrodynamics simulations \cite{Lockitch:2002sy,Idrisy:2014qca}. It depends in addition to the frequency on the equation of state and the particular star configuration which can be parametrized by its compactness parameter $M/R$. Whereas initial analyses \cite{Lockitch:2002sy} considered simple equations of state, like polytropes, a recent analysis took into account realistic nuclear matter equations of state \cite{Idrisy:2014qca} and showed that the results are very insensitive over the wide range of equations of state considered, reflecting that the r-mode is a dominantly non-radial mode. This striking result allows us to extract quantitative information from future gravitational wave data, despite the significant uncertainties in the description of both the micro- and the macrophysics. 

\subsection{Multipole analysis of gravitational wave emission}
Since compact stars are perfect point sources, their gravitational wave emission can be described within a multipole expansion \cite{Thorne:1980ru}. Here we give the expressions for the r-mode current quadrupole moment and the relevant derived expressions \cite{Lindblom:1998wf,Owen:1998xg} in a slightly generalized form that takes into account a general r-mode spectrum parametrized by the function $\chi(\Omega)$ in eq.~(\ref{eq:gw-frequency}).

The perturbation of a flat background Minkowski metric due to current quadrupole radiation of a source at the origin is given by \cite{Thorne:1980ru}
\begin{equation}
h_{ij}^{TT}=\frac{\sqrt{G} }{r} \frac{d^2}{(dt)^2} S_{22}\!\left(t-r\right) T_{22,jk}^{B2}
\end{equation}
in terms of the quadrupolar tensor spherical harmonics $T_{22}^{B2}$. The current quadrupole moment has the general form
\begin{equation}
S_{22}=-\frac{32\sqrt{2}\pi}{15}\int \tau_{0j} Y_{22,j}^{B*}r^{l}d^{3}x
\end{equation}
where $\tau$ is the effective stress-energy tensor \cite{Thorne:1980ru}, and its relevant off-diagonal component reads in the non-relativistic limit $\tau_{0j} \!=\! \rho\,\delta v_j$. For the non-relativistic form of the r-mode eq.~(\ref{eq:r-mode}) this gives
\begin{equation}
S_{22}=-\frac{32\sqrt{2}\pi}{15}\tilde{J} \alpha MR^{3}\Omega e^{i\omega t} \label{eq:quadrupole}
\end{equation}
where the dimensionless constant $\tilde J$ specifies the radial density profile\footnote{This expression holds for a spherically symmetric star. Rotation only leaves axial symmetry and therefore in general a slightly more complicated expression for $\tilde J$ has to be employed, which would have quantitative impact for very fast rotating sources.}
\begin{equation}
\tilde{J} \equiv\frac{1}{MR^2}\int \! \rho \, f\!\left(\frac{r}{R}\right)r^4 d r \label{eq:J-tilde}
\end{equation}

The detectability of gravitational waves in a terrestrial detector is standardly described in terms of the intrinsic gravitational wave strain amplitude\footnote{It represents the amplitude measured in a hypothetical detector placed at a pole from a source vertically above the detector whose rotation axis is aligned with that of the earth.} $h_0$. In the case of r-modes it is obtained from the quadrupole expression eq.~(\ref{eq:quadrupole})
\begin{equation}
h_{0}=\sqrt{\frac{2^{15}\pi G}{3^{6}5}}\tilde{J}MR^{3} \frac{\chi^{2} \Omega^{3}\alpha}{D} \label{eq: intrinsic-strain}
\end{equation}

In addition to the direct gravitational wave signal several energy change rates are of interest for the r-mode analysis. The power pumped into the unstable r-mode by gravitational radiation reaction is given by eq.~(7) in \cite{Lindblom:1998wf} and yields for a general frequency relation eq.~(\ref{eq:gw-frequency})
\begin{equation}
P_G \equiv \frac{dE_{\rm mode}}{dt}=\hat G \left(1+3\frac{1-\chi}{\chi}\right)\chi^{6}\alpha^{2}\Omega^{8} \label{eq:instability-power}
\end{equation}
generalizing the corresponding result in \cite{Lindblom:1998wf}. Here the constant $\hat G$ is determined by static source properties
\begin{equation}
\hat G\equiv \frac{2^{17}\pi}{3^{8}5^{2}} \tilde{J}^{2} G M^{2}R^{6}
\end{equation}
The gravitational wave luminosity is given by eq.~(4.16) in \cite{Thorne:1980ru} and yields
\begin{equation}
L_G\equiv\frac{dE_{\rm wave}}{dt}=2 \hat G \chi^6 \alpha^2 \Omega^8 \label{eq:gw-luminosity}
\end{equation}
The rotational energy loss rate due to r-modes is then
\begin{equation}
P_R \equiv \left.\frac{dE_{\rm rot}}{dt}\right|_{R} = -\left(P_G+L_G\right) = -3\hat G \chi^{5} \alpha^{2}\Omega^{8} \label{eq:rotational-loss}
\end{equation}
I.e. $2\chi/3$ of the lost rotational energy is radiated off as gravitational waves and the residual about $1/3$ is pumped into the mode by the instability and is eventually dissipated to heat within the star.

Using eq.~(\ref{eq: intrinsic-strain}) the r-mode amplitude can be expressed in terms of the intrinsic strain amplitude that would be obtained from a gravitational wave detection
\begin{equation}
\alpha=\frac{2}{3\sqrt{5\hat{G}}}\frac{Dh_{0}}{\chi^{2}\Omega^{3}} \label{eq:r-mode-alplitude}
\end{equation}
in order to express the above powers in terms of the intrinsic strain amplitude \cite{Owen:2010ng}. This gives for the various energy change rates 
\begin{eqnarray}
L_G&=&\frac{8}{45}\chi^2\Omega^{2}D^{2}h_{0}^{2} \label{eq:gw-luminosity} \\
P_{G}&=&\frac{4}{45}\left(1+3\frac{1-\chi}{\chi}\right)\chi^{2}\Omega^{2}D^{2}h_{0}^{2} \label{eq:-r-mode-heating} \\
P_{R}&=&\frac{4}{15}\chi\Omega^{2}D^{2}h_{0}^{2} \label{eq:r-mode-spindown-power}
\end{eqnarray}
Note that these expressions are independent of the generally unknown source properties contained in the constant $\hat G$.

\section{R-mode gravitational wave scenarios}
\label{sec:scenarios}

\begin{figure*}
\resizebox{0.33\textwidth}{!}{
\includegraphics{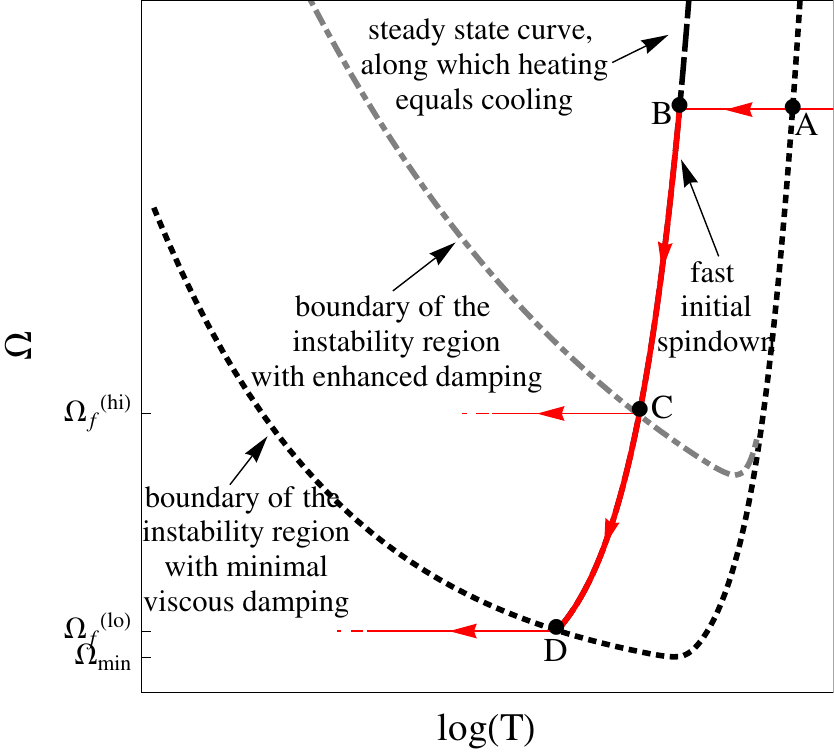}
}
\resizebox{0.33\textwidth}{!}{
\includegraphics{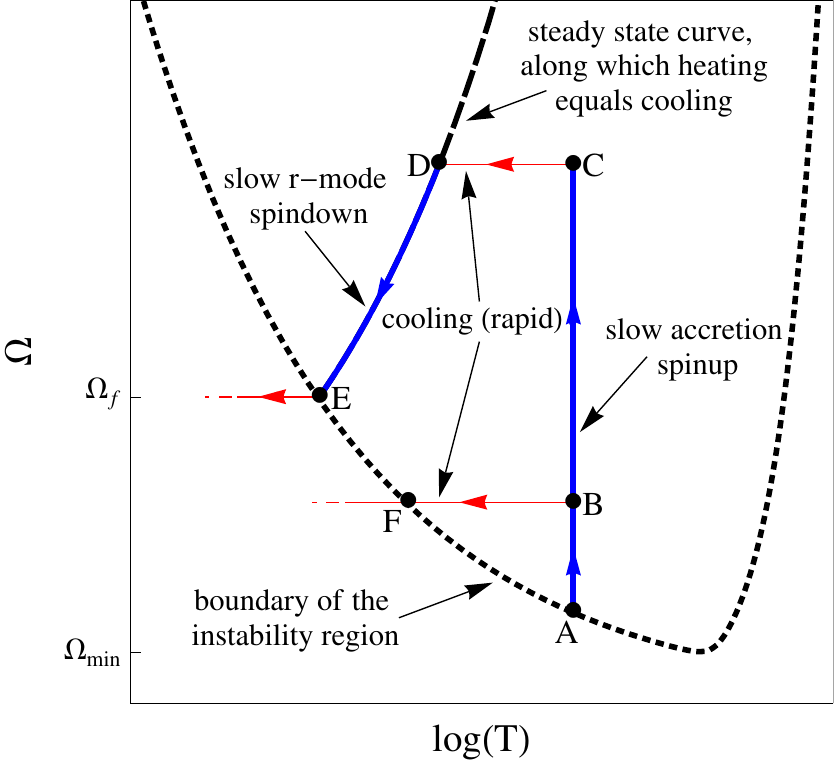}
}
\resizebox{0.33\textwidth}{!}{
\includegraphics{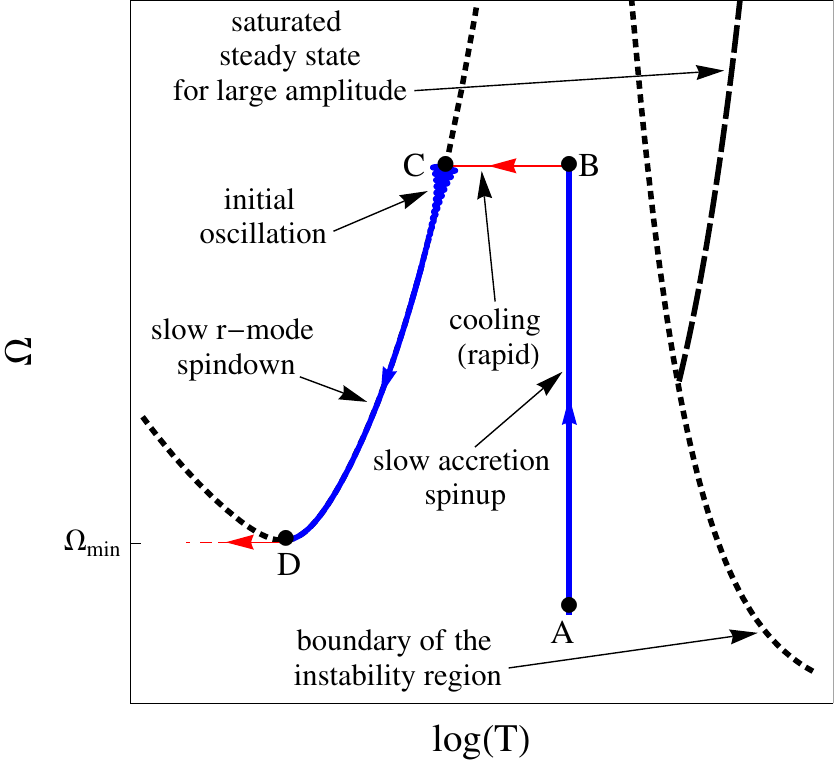}
}
\caption{Exemplary r-mode gravitational wave emission scenarios for sources at different stages in their evolution. {\em Left}: Initial spindown of newborn and young sources; {\em middle}: Recycling in LMXBs and spindown of millisecond pulsars in the saturated r-mode scenario; {\em right}: Recycling and spindown in the no r-mode and boundary-straddling \cite{Andersson:2001ev} scenarios realized for sources with enhanced damping imposing a stability window. Whenever the evolution runs inside or at the boundary of the instability region (bounded by the dotted lines) there is gravitational wave emission. See the text for details.}
\label{fig:scenarios}
\end{figure*}

The r-mode emission from compact sources is significantly more diverse than the emission due to ellipticity since it strongly depends on the evolution of a source and thereby on its age and its internal composition. In this section we discuss the relevant classes of sources and the different scenarios within which r-mode gravitational emission could be present. R-modes become only unstable in sources that spin sufficiently fast since viscous dissipation can otherwise damp them away. Since the damping is strongly temperature dependent r-modes are unstable in characteristic instability regions in a $T$-$\Omega$-diagram \cite{Andersson:1998ze,Andersson:2000mf,Alford:2010fd} and a few possibilities, depending on the composition of the star, are shown in fig.~\ref{fig:scenarios}. The largest instability region is obtained for standard neutron stars with well established viscous damping mechanisms \cite{Shternin:2008es,Sawyer:1989dp,Friman:1978zq} (``minimal damping''), but potential enhanced damping, e.g. in certain exotic forms of matter, could reduce it (left panel) or induce a stability window where the r-mode is stable up to large frequencies (right panel) \cite{Madsen:1998qb,Alford:2010fd}.

It is known that pulsars have a varied evolution: whereas newly-formed compact stars are expected to be born with large spin frequencies due to the dramatic spin-up during core collapse, the observed young pulsars all have rather low spin frequencies but still large spindown rates. Generally, during their evolution they spin down further and their radio emission stops once their spindown power becomes too small to power the pulsar jets. However, when compact stars happen to be in a binary they can be recycled by accretion spin-up, and being heated by nuclear reactions they emit thermal X-ray radiation. Sources in such low mass X-ray binaries (LMXBs) reach frequencies of many hundreds of Hertz (see fig.~\ref{fig:LMXB-limits}). Once the accretion stops they are not strongly heated any more and turn into stable millisecond pulsars with very precise observed timing data and which hardly spin down. Based on this general evolution there are four classes of sources in which r-modes could be unstable, namely 

\begin{enumerate}
\item newly formed (proto-)compact stars in the immediate aftermath of a supernova explosion or merger event
\item young sources, under a thousand years old, that are still in their initial rapid spindown
\item sources in low mass X-ray binaries (LMXBs) that accrete and on average spin up
\item millisecond pulsars that can have ages of billions of years and only very slowly spin down
\end{enumerate}

In case of ellipticity-based gravitational wave emission these classes of sources are most promising as well since the gravitational wave strain scales strongly with the emission frequency, but emission is in this case in principle expected in all rotating sources. In contrast r-modes cannot be present in all other sources, like the observed young pulsars as well as the many other sources which spin with frequencies significantly below 100 Hz, since r-modes are viscously damped in these sources \cite{Alford:2010fd,Alford:2012yn}. R-modes could also be temporarily triggered by transient events like star quakes or burst, but we will not discuss such transient cases here, since they should be harder to detect. 

Whether r-modes are actually present in any of the above sources depends strongly on the current evolutionary state of the star and the dissipation within its interior. To understand within which scenarios r-modes can be present in the above sources requires to study the compact star evolution in the presence of r-modes, described by the r-mode amplitude $\alpha$, the star's rotational angular momentum $\Omega$ and its temperature $T$. The evolution is determined by global energy and angular momentum conservation \cite{Owen:1998xg,Alford:2012yn} within the star and driven by the powers in eqs.~(\ref{eq:gw-luminosity}-\ref{eq:r-mode-spindown-power}). As discussed in \cite{Alford:2012yn} the evolution of the different quantities happens on very different time scales and therefore effectively decouples. If not damped the r-mode amplitude rises quickly on a time scale of order seconds at millisecond frequencies \cite{Owen:1998xg,Kokkotas:1998qr,Andersson:2000mf}. In case the source stays in the instability region, non-linear effects eventually strongly enhance the damping at large amplitude and saturate it at a finite value $\alpha_{\rm sat}$. 
At this point it is not known which of the proposed mechanisms \cite{Rezzolla:1999he,Lindblom:2000az,Arras:2002dw,Cuofano:2009yg,Alford:2011pi,Haskell:2013hja,Alford:2014jha} stops the growth and saturates the mode. R-mode saturation amplitudes generally depend on the angular velocity and temperature of the star and have a power law dependence $\alpha_{\rm sat}=\hat\alpha_{\rm sat} T^\beta \Omega^\gamma$, where the parameters $\beta$ and $\gamma$ describe the so far unknown saturation mechanism. It has been shown \cite{Alford:2012yn} that during the subsequent saturated stage the thermal evolution is always faster than the spindown so that the star tends towards a thermal steady state where the heating balances the cooling. If this steady state is reached the spin evolution evolves along the corresponding steady state curve \cite{Alford:2013pma}. In general the heating depends strongly on the r-mode saturation amplitude and therefore at a lower saturation amplitude this spindown curve runs at lower temperatures. Another generic feature of the evolution is that both the cooling rate and the spindown rate depend strongly on temperature and frequency and become small as they decrease, so that the evolution slows down.

\subsection{Newly-formed and young compact stars}

The evolution of newly-formed and young sources \cite{Lindblom:1998wf} is shown on the left panel of fig.~\ref{fig:scenarios} (solid curve). A newly-formed source is very hot and should rotate at a frequency close to its Kepler limit. In case the hot source is initially in the r-mode stable regime, as would be the case for minimal viscous damping in neutron stars \cite{Lindblom:1998wf,Alford:2010fd,Alford:2012yn}, it will cool until it reaches at some point (A) the instability region (dotted) where the amplitude grows until it is saturated by nonlinear effects. If the saturation amplitude is high the very fast initial spindown can lead to a prompt r-mode gravitational wave signal in newly-formed sources that is very strong for a short time interval of the order of minutes \cite{Owen:1998xg,Mytidis:2015}. After a rapid initial cooling phase young sources reach their thermal steady state (B), where r-mode heating matches radiative cooling, and then spin down along the steady state curve (dashed), emitting continuous gravitational radiation. Depending on the saturation amplitude this can last for up to several hundred years. When the evolution reaches the boundary of the instability region (D) the r-mode spindown ends at a frequency $\Omega_f^{({\rm lo})}$ (slightly above the minimum of the instability region $\Omega_{\rm min}$) where the r-mode decays and the gravitational wave emission stops. For neutron stars with minimal viscous damping this happens below 100 Hz \cite{Alford:2012yn}. However, when enhanced damping is present in compact stars, as illustrated by the lighter (dot-dashed) instability region, r-modes could already decay at much higher frequencies $\Omega_f^{({\rm hi})}$ (C) and the gravitational wave emission would correspondingly be stopped much earlier.

\subsection{Sources in low mass X-ray binaries}

Stars being recycled in LMXBs are strongly heated by accretion and potentially r-modes. Known sources in LMXBs have temperatures of order $10^8$ K \cite{Haskell:2012} as obtained from thermal X-ray fitting \cite{Ozel:2012wu}. This places them well inside the instability region of neutron matter with minimal damping \cite{Andersson:2000pt,Haskell:2012,Alford:2013pma} and this statement is not affected\footnote{Due to the smallness of the scale ratio $T/\mu\lesssim 10^{-4}$ in compact stars the boundary of the instability region depends strongly on the exponents of the generic power law temperature dependence of the viscosities, but hardly on their quantitative prefactors, that are poorly constraint due to uncertainties in the microphysics \cite{Alford:2010fd}.} by the various uncertainties \cite{Alford:2014nba}. As shown on the middle panel of fig.~\ref{fig:scenarios}, while spinning up such sources would in the minimal damping scenario enter the r-mode instability region from below, where the r-mode would again be saturated by some non-linear mechanism. In this {\em saturated r-mode} scenario the star would steadily emit gravitational waves as the source spins up and this could last for many millions of years. Depending on the relative strength of accretion and r-mode heating, the spin-up could either proceed along a vertical line (from A to C) being permanently heated by accretion, or along the r-mode steady state curve (from E to D) in case r-mode heating dominates\footnote{The latter case would require that the observed LMXB temperatures (see e.g. fig.~2 in \cite{Alford:2013pma}) all lie on a single monotonic thermal steady-state curve. Although the present data strongly scatter, this cannot be ruled out due to the significant uncertainties in the present temperature estimates. Yet, this might be challenged by more precise data. However, if accretion heating is dominant at least in the hotter sources, this could easily explain the scattered observational data, since different sources have different accretion rates.}. In any case the temperature measurements in LMXBs set very stringent bounds on the r-mode saturation amplitude in these sources $\alpha \lesssim 10^{-8}-10^{-7}$ \cite{Mahmoodifar:2013quw,Alford:2013pma}. Current saturation mechanisms face a severe challenge to provide such low saturation amplitudes, as would be required in standard neutron stars without enhanced damping mechanisms \cite{Alford:2013pma}. These bounds also exclude previously proposed cyclic evolution scenarios\footnote{In these scenarios relying on a large saturation amplitude $\alpha \ggg 10^{-7}$, sources would periodically experience a long cooling and spin-up epoch, followed by a quick heating and spindown phase during which they would emit copious gravitational waves. The many sources within the instability region with much lower temperatures contradict such a scenario.} that relied on a large saturation amplitude \cite{Levin:1999ApJ...517..328L,Heyl:2002pe}. It has also been proposed \cite{Andersson:2000pt} that r-modes could explain the spin frequencies of LMXBs and millisecond pulsars which spin significantly below their theoretical Kepler limit, although there are also other potential explanations for this finding, see e.g. \cite{White:1997,Bildsten:1998ApJ}. The spin limit would arise from the r-mode spindown torque balancing the accretion torque, but the bounds on the r-mode amplitude also challenge this scenario.  

In case enhanced damping due to certain exotic forms of matter, leading to a stability window, is present, sources in LMXBs, heated by accretion, can be spun up to large frequencies without entering the r-mode instability region, as shown on the right panel of fig.~\ref{fig:scenarios}. In this {\em no r-mode} scenario there would be no r-mode gravitational wave emission from LMXB sources (vertical line from A to B). An alternative possibility is that nuclear heating is weaker and the stability window is located at higher temperatures. In this {\em boundary-straddling} scenario a non-linear mechanism that saturates the r-mode at a low amplitude is not required and the saturation value can be very large. Instead the r-mode amplitude is dynamically kept at a low amplitude since the strong r-mode heating periodically pushes a source out of the instability region where the r-mode decays and the source cools in again \cite{Andersson:2001ev}. This results eventually in a dynamic equilibrium with a boundary-straddling amplitude and sources would spin up along the lower boundary of the stability window (curve bounded by D and C) and would correspondingly emit gravitational waves\footnote{This scenario would again require the LMXB temperature data to lie on a monotonous curve.}. 

\subsection{Millisecond pulsars}

Old millisecond pulsars originate from LMXBs after the accretion stops and correspondingly it depends on their previous evolution if such sources emit gravitational waves. In case the minimal damping scenario, valid for standard neutron matter, is realized, the sources are within the instability region when the accretion stops as shown in the middle panel of fig.~\ref{fig:scenarios}. In case during the LMXB phase accretion heating dominated, a fast spinning source will first quickly cool until it reaches the thermal steady state curve (horizontal line from C to D). Independent of the evolution history, millisecond pulsars with saturated r-modes will then very slowly spin down along the steady state curve (curve from D to E). At the tiny spindown rates of observed sources they cannot leave the instability region and would in the saturated r-mode scenario emit gravitational waves for billions of years. As discussed in  \cite{Alford:2013pma}, the only way that a fast spinning source could leave the instability region is that the r-mode saturation amplitude is very small $\alpha<10^{-10}$ in which case it could cool without being stopped by the r-mode heating. However, none of the presently proposed saturation mechanisms can provide such a low saturation amplitude. Slowly spinning millisecond pulsars with frequencies below 200-300 Hz, however, can quickly cool out of the instability region (horizontal segment from B to F) so that gravitational waves are not expected from such sources \cite{Alford:2014pxa}.

If enhanced damping, e.g. due to certain exotic forms of matter, is present that leads to a stability window the right panel of fig.~\ref{fig:scenarios} applies again. In this case a source heated by accretion in LMXBs would after the accretion stops likewise cool (horizontal segment from B to C) and then in the boundary straddling scenario bounce around the boundary of the instability window \cite{Andersson:2001ev}. After the initial oscillation has decayed a dynamic equilibrium is reached with a dynamic r-mode amplitude and a source then slowly spins down along the boundary of the instability region (curve from and C to D) and likewise emits gravitational waves for billions of years. Correspondingly, both ordinary and exotic forms of matter allow for gravitational wave emission from fast spinning millisecond pulsars which would present very interesting sources for r-mode astronomy due to their extreme stability allowing for precision measurements.

\section{Detectability}
\label{sec:detectability}

For realistic estimates on the chances to use the continuous r-mode emission from compact sources for gravitational wave astronomy, it is crucial to determine the strength of these signals \cite{Owen:1998xg,Bondarescu:2008qx,Owen:2010ng,Alford:2012yn,Mahmoodifar:2013quw,Alford:2014pxa,Mytidis:2015} for the different classes of sources. The detectability of gravitational waves from r-modes obviously depends on the gravitational wave amplitude which according to eq.~(\ref{eq: intrinsic-strain})
is directly proportional to the r-mode amplitude $\alpha$ and also depends on the frequency and the distance of a given source.

\subsection{Newly-formed compact stars}

In the aftermath of a supernova explosion, large amplitude r-modes could quickly spin-down a rapidly rotating star and correspondingly this would result in an intense but short initial peak in the gravitational wave signal. The gravitational wave emission from newly-formed sources had initially been studied in \cite{Owen:1998xg} using a frequency- and temperature-independent toy model for the saturation mechanism. A generic result was that although the initial power is very large the overall gravitational wave energy in this transient spike is rather small compared to the subsequent continuous emission. In \cite{Bondarescu:2008qx} the gravitational wave emission had been studied for a more realistic mode-coupling saturation mechanism, yet in this case exotic processes like hyperon bulk viscosity had to be invoked in order to obtain proper saturation for the considered mechanism.
Recently the case of newly-formed sources has been re-analyzed in \cite{Mytidis:2015} where the authors performed a detailed study in order to determine the distance up to which a source could be detectable with advanced LIGO \cite{TheLIGOScientific:2014jea} and the planned Einstein telescope \cite{Punturo:2010zz}. This study points out that for a r-mode amplitude $\alpha= O(1)$ we probably should have already detected a signal from recent supernovae with previous detectors\footnote{Such large amplitudes are also incompatible with the ages of the fastest young pulsars \cite{Alford:2012yn}.}. Relying on the saturation model of \cite{Owen:1998xg} with $\alpha=0.1$, this study finds that with next generation detectors newly-formed sources resulting from a supernova can in the best case be detected up to a distance of 1 Mpc, i.e. within our local group of galaxies. The estimated rates for such events are however not high enough that a detection is very likely with second generation detectors.  

Newly-formed compact stars will likely first be detected by a direct gravitational wave signal from the supernova or merger, the neutrino burst or the subsequent electromagnetic signal so that the position will be known, just as for the other classes studied below. However spin information is not available so that the data analysis for the source is more involved and requires a large set of search templates. This complexity is only partly compensated by the large signal strength and correspondingly the advantage to restrict the search to a shorter data set than for the other searches discussed below. A currently unaddressed problem for a search for newly-formed sources could be that in the immediate aftermath of a supernova the star is still not sufficiently equilibrated that the generic r-mode templates that are currently used \cite{Owen:1998xg,Mytidis:2015} are sufficiently close to the actual signal due to various complications, like other oscillation modes that are not yet damped, accretion onto the proto star, or the ongoing differentiation in its interior. 

An interesting possibility is that during and after the core bounce r-modes are externally triggered at high amplitude. Both observations \cite{Boggs:2015Sci} and recent simulations, see e.g. \cite{Yakunin:2015wra}, show that supernovae are generically asymmetric. In the simulations this stems from strong convection in the compact core that could excite non-radial modes, like r-modes. In this case the instability of r-modes and their saturation amplitude would be less important and an appreciable gravitational wave signal could be possible even if the saturation amplitude is actually low, since it would take time for a excited large amplitude mode to decay to its saturation level.

\subsection{Young sources}

Whereas newly formed sources are expected to spin with large frequencies close to their Kepler limit, observed young pulsars which are all older than about a thousand years spin very slow, $f \lesssim 62$ Hz. Therefore, some mechanism must spin down young sources during this time and r-modes are an interesting possibility. Taking into account that all known young pulsars spin already too slow for r-modes to be present, the class of young sources consists of stars that we either haven't observed at all, yet, or for which we at least haven't observed pulsation and correspondingly do not know their rotation frequency. These are either recent supernova remnants without an associated compact object, like SN1987A, or young neutron stars like Cassiopeia A. This class is different from the other source classes since there is a convincing indication that very young sources could indeed spin down by r--modes and therefore actually show gravitational wave emission. The reason for this is that r-modes strikingly provide a {\em quantitative} explanation for the low spin frequencies of young pulsars \cite{Alford:2012yn}, which are more than an order of magnitude below their Kepler frequency limit. Although magnetic emission can likewise spin-down young sources this mechanism does not impose an upper frequency limit and therefore it does not explain why there is not a single source that spins with higher frequencies.

\begin{figure}
\resizebox{0.5\textwidth}{!}{
\includegraphics{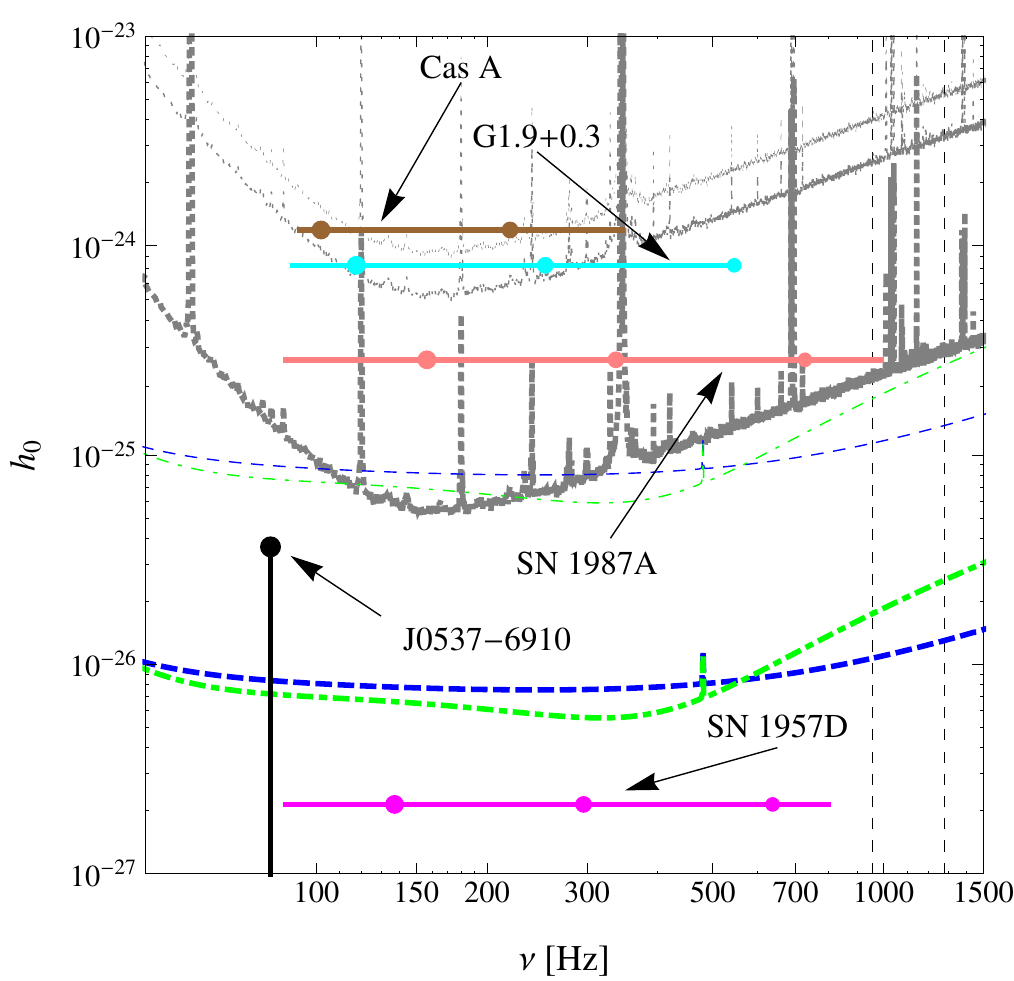}
}
\caption{Predictions for the intrinsic
gravitational wave strain amplitude $h_0$ for potential young sources due to r-modes \cite{Alford:2012yn}. The actual age of
these sources as well as their distance have been used and the expected
frequency range determined from these known parameters is shown. The
frequency range corresponds to a range of saturation amplitudes and
the dots denote the corresponding values for different orders of magnitude
from $\alpha_{{\rm sat}}=0.1$ to $10^{-3}$. We also included the
fastest young pulsar J0537-6910 for which the observed frequency is
used and where the dot represents the result if r-mode spindown dominates
and lower values are possible if other spindown sources are important.
The theoretical results are compared to the sensitivity of LIGO (solid) and the anticipated
sensitivity of advanced LIGO \cite{Harry:2010zz} in the standard
mode (dashed) and the neutron star optimized configuration (dot dashed).
These are given both for a known pulsar search with one year of data
(thick), a search for potential sources without timing information
using one month of data (thin), as well as the sensitivity
of a previous Cas A search \cite{Abadie:2010hv} (uppermost, very thin curve).
The two thin vertical
lines show for comparison the corresponding frequency of the fastest
observed (old) pulsar and that corresponding to the maximum Kepler
frequency for the considered model of a compact star. }
\label{fig:strain-sources}
\end{figure}

To provide an explanation for the rapid spindown of young pulsars, requires a sufficiently large r-mode saturation amplitude $\alpha_{\rm sat} \gtrsim 10^{-3}$ and for such amplitudes the r-mode gravitational wave signal is {\em independent} of the saturation amplitude \cite{Wette:2008hg} if the initial spin-frequency $\Omega_i$ was sufficiently above the current frequency $\Omega$  \cite{Alford:2012yn} and takes the simple form
\begin{equation}
h_0\!\left(t\right) 
= \sqrt{\frac{5}{8}\frac{{\cal C}GI}{D^{2}t}}\:.\label{eq:independent-strain}
\end{equation}
Here $D$ is the distance to the source, $t$ its age, $I$ its moment of inertia and
\begin{equation}
{\cal C}\equiv\frac{1-2\beta/\theta}{1+\gamma/3+2\beta/\left(3\theta\right)}
\end{equation}
is a constant that encodes the saturation and cooling mechanism and varies for the proposed mechanism \cite{Rezzolla:1999he,Lindblom:2000az,Arras:2002dw,Cuofano:2009yg,Alford:2011pi,Haskell:2013hja,Alford:2014jha} moderately from its value ${\cal C}=1$ for a constant saturation amplitude. This scenario yields therefore, in contrast to the others discussed in this section, a {\em definite prediction} for the r-mode gravitational wave strain that depends mainly on the age and distance of a given source \cite{Alford:2012yn}.
Taking into account the uncertainties on the parameters gives the prediction
\begin{equation}
h_{0} \approx2.3_{-0.8}^{+3.5}\times10^{-27}\sqrt{\frac{1000\,{\rm y}}{t}}\frac{1\,{\rm Mpc}}{D}\:,\label{eq:numeric-intrinsic-strain}
\end{equation}
where the uncertainties stem from our ignorance on global source properties. Fig.~\ref{fig:strain-sources} shows the signal for various promising sources compared to the sensitivity of the LIGO and aLIGO detectors. It shows that whereas it is perfectly consistent that LIGO did not see a signal, several sources should be significantly above the sensitivity of aLIGO if r-modes are indeed responsible for the spindown of young pulsars.

Whether there is gravitational wave emission with the strength eq. (\ref{eq:numeric-intrinsic-strain}) depends on the dissipation within the star. For neutron stars with minimal viscous damping emission is possible for hundreds of years and from sources spinning with frequencies down to below 100 Hz. Several such known sources are shown in fig.~\ref{fig:strain-sources} as well as the fastest young pulsar J0537-6910 for which it cannot be completely excluded that it is still in the last stage of its r-mode spindown phase \cite{Alford:2012yn}. For the sources without spin information, depending on the unknown saturation amplitude, the given frequency ranges shown in fig.~\ref{fig:strain-sources} are possible. When enhanced damping is present in a source this would reduce the time interval within which gravitational waves are emitted as well as the frequency range. If the saturation amplitude is too high r-modes could then already be damped in the observed sources. However, this way r-modes would fail to provide the striking quantitative explanation for the low spin frequencies of the observed young pulsars.

\subsection{Sources in LMXBs}

Sources in LMXBs intermittently accrete matter from their companion star and therefore on average spin up. As discussed above r-mode gravitational wave emission is expected for neutron stars with standard damping, but should be absent for certain exotic forms of matter. The detectability of continuous gravitational waves due to r-modes in LMXBs has previously been discussed in \cite{Watts:2008qw}, where the emission was discussed under the assumption that the accretion in these systems is balanced by r-mode spindown. This may be the case but is not substantiated by astrophysical observations. Instead general bounds on the gravitational wave emission have recently been obtained \cite{Mahmoodifar:2013quw} from the thermal steady state that is realized when r-modes are saturated \cite{Alford:2012yn}. These bounds stem from the observation that large-amplitude r-modes would strongly heat these sources and correspondingly very stringent bounds on both the r-mode amplitude and the gravitational wave strain can be placed \cite{Mahmoodifar:2013quw}. Taking into account that in LMXBs there is additional heating due to the accretion, in the thermal steady state the r-mode heating eq.~(\ref{eq:-r-mode-heating}) alone cannot be larger than the cooling in the star $P_G \leq L_\gamma + L_\nu$, where $L_\gamma$ and $L_\nu$ are the neutrino and photon luminosity which both scale with the temperature via a power law $L_i = \hat L_i T^{\theta_i}$ \cite{Alford:2013pma}. With observational temperature estimates from thermal fits to the X-ray spectra this can be resolved to obtain bounds on the intrinsic gravitational strain amplitude, as done in \cite{Mahmoodifar:2013quw}. This work finds that due to the thermal constraints the signal is significantly below the aLIGO sensitivity threshold. These bounds are only reached if r-modes dominate the heating of these sources and can even be significantly lower if heating from accretion and processes in the crust \cite{Brown:1998ch} determine the observed temperatures of these sources. 

\begin{figure}
\includegraphics{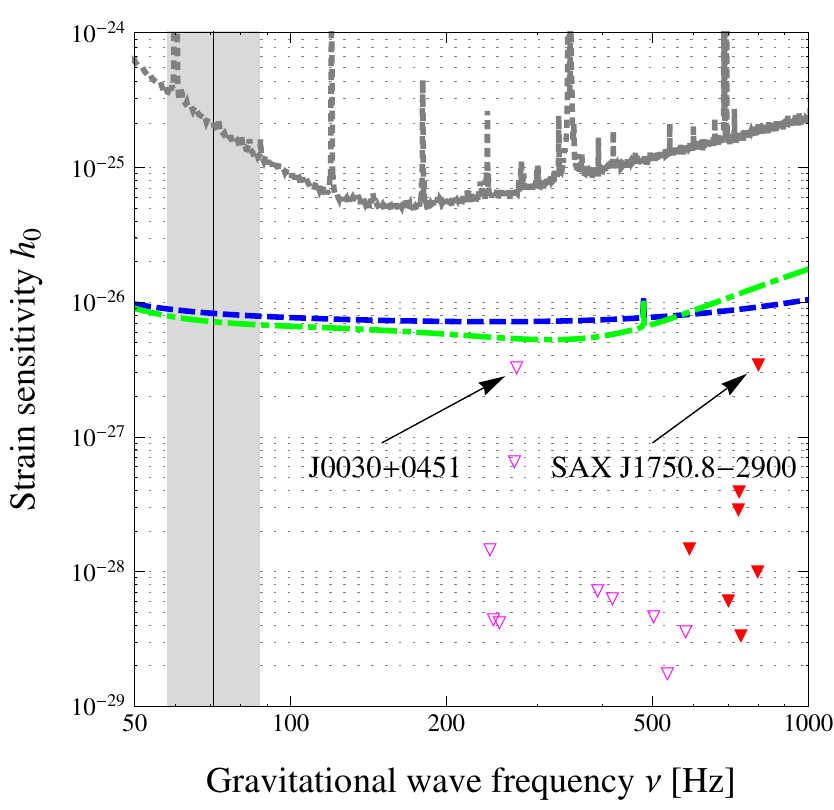}
\caption{Upper bounds on the intrinsic strain amplitude for sources in LMXBs obtained from their temperature estimates. Solid triangles show sources with temperature measurements whereas open triangles show sources for which only upper temperature bounds are present. Whereas the former are inside of a r-mode instability region for a neutron stars with minimal damping, the latter could be cold enough that they are outside and correspondingly there would be no gravitational wave emission at all.}
\label{fig:LMXB-limits}
\end{figure}

We reanalyze the gravitational wave detectability from LMXBs in fig.~\ref{fig:LMXB-limits} using thermal data given in \cite{Haskell:2012} and compare the bounds on the intrinsic gravitational wave strain again to the detector sensitivity of LIGO and aLIGO. We also include a few additional sources that were not considered in \cite{Mahmoodifar:2013quw}. Although we generally confirm the previous results, we find the weakest bounds, and therefore the best chance for a detection, for sources that were not included in \cite{Mahmoodifar:2013quw}. These sources, like SAX J1750.8-2900 are about an order of magnitude above those discussed previously, so that even though they are below the estimated sensitivity of aLIGO, a coherent analysis of data spanning more than a year, further detector enhancements or third generation detectors could provide a realistic chance to detect these sources.

\subsection{Millisecond pulsars}

Whether there can be r-modes in old isolated millisecond radio or high-energy pulsars depends strongly on the evolution of these sources, as discussed in sec.~\ref{sec:scenarios}. Millisecond pulsars are so old that they would easily have cooled out of the unstable zone unless internal heating prevents this. In millisecond pulsars the dominant heating source stems from r-mode oscillations themselves and it depends on the frequency of the source if this heating can prevent millisecond pulsars from cooling. A detailed analysis shows that r-modes should be absent in sources spinning slower than $f \lesssim 200$ Hz \cite{Alford:2014pxa}. A standard bound on the gravitational wave emission of these source comes from the {\em spindown limit} \cite{Owen:2010ng} which relies on the fact that the r-mode spindown power eq.~(\ref{eq:r-mode-spindown-power}) is constrained by the observed frequency and spindown rate (timing data) of a given source $P_R \leq I \Omega \dot \Omega$. Resolved for the gravitational wave strain this yields an upper limit for it
\begin{equation}
h_{0}^{\left({\rm sl}\right)}=\sqrt{\frac{15}{4}\frac{GI\left|\dot{f}\right|}{D^{2} f}}\:,
\end{equation}

Recently, it has been shown that an even stronger limit stems from the constraint that the same r-mode saturation mechanism should be active in all observed sources \cite{Alford:2014pxa}. In this case a given source is constrained by the timing data of the entire set of millisecond pulsars which yields the stronger {\em universal spindown limit} \cite{Alford:2014pxa} on the gravitational wave strain of observed sources in the saturated r-mode scenario 
\begin{equation}
h_{0}^{\left({\rm usl}\right)}=\sqrt{\frac{15}{4}\frac{GI\left|\dot{f}_{0}\right|}{D^{2}f_{0}}}\left(\frac{\hat{\alpha}_{{\rm sat}}^{\left({\rm mac}\right)}}{\hat{\alpha}_{{\rm sat},0}^{\left({\rm mac}\right)}}\right)^{{\textstyle \frac{1}{1-2\beta/\theta}}}\left(\frac{f}{f_{0}}\right)^{{\textstyle \frac{3+\gamma+2\beta/\theta}{1-2\beta/\theta}}}\:.
\end{equation}
Here the quantities labeled by a subscript 0 refer to the source whose timing data sets the most stringent constraint on the r-mode saturation amplitude and which can depend on the considered saturation mechanism.

\begin{figure}
\includegraphics{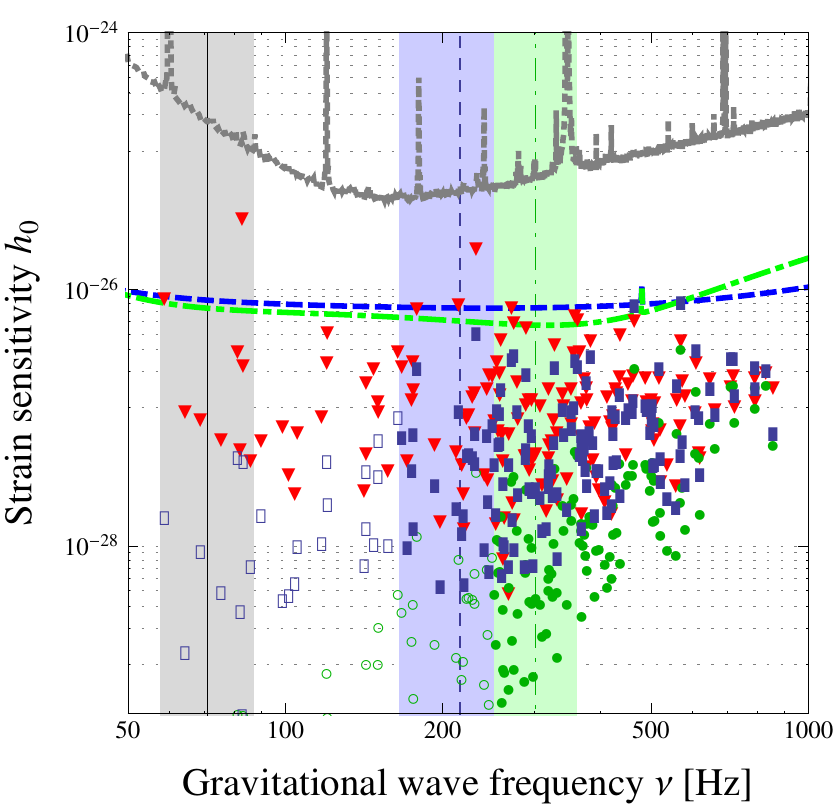}
\caption{Comparison of different upper
bounds on the strain amplitude of known radio pulsars due to r-mode
emission \cite{Alford:2014pxa}. The spindown limit (red triangles) is obtained from the
timing data of an individual source. The universal spindown limit
takes into account that the saturation mechanism applies to the entire
class of millisecond pulsars, and also provides a lower bound 
on the frequency below which gravitational wave r-mode emission is excluded (see sec.~\ref{sec:scenarios}). 
We show universal spindown limits (green circles)
and minimum frequency (green dot-dashed vertical line with uncertainty
band) for the toy model of a constant r-mode saturation amplitude
\cite{Owen:1998xg}. We also show universal spindown limits (blue
rectangles) and minimum frequency (blue dashed vertical line with
uncertainty band) for a realistic saturation mechanism arising from
mode-coupling and the damping of the daughter modes by shear viscosity
\cite{Bondarescu:2013xwa}. For a given saturation mechanism, stars
below the minimum frequency (open symbols) do not undergo r-mode oscillation.}
\label{fig:universal-spindown-limits}
\end{figure}

These limits are shown for the observed millisecond pulsars in fig.~\ref{fig:universal-spindown-limits}. As can be seen the ordinary spindown limits (triangles) of some sources are above the detection sensitivity of aLIGO. In contrast, the universal spindown limits, shown for two different saturation mechanisms (rectangles and circles), are below the detection limit. In particular, it is the fast spinning sources that are most promising and several of them are very close so that future enhancements could likewise give us the chance to detect these source. For the alternative boundary-straddling scenario, realized when certain exotic forms of matter are present in these sources, analogous limits have been derived and yield quantitatively very similar results \cite{Alford:2014pxa}. 

\section{Identification and direct observables}
\label{sec:identification}

\subsection{Identification}

In case the rotation frequency of the source is known, the observed gravitational wave frequency can directly confirm that the emission stems from r-modes. The two important classes of continuous gravitational wave emission from compact stars are deformations (ellipticity) and r-modes. In case of elliptic deformations the gravitational wave frequency is exactly twice the rotation frequency $\nu=2 f$ which would be a clear signature. For r-modes the gravitational wave frequency equals the frequency of the mode in the inertial frame. In Newtonian slow-rotation approximation the canonical relation for the fundamental $l=m=2$ r-mode is $\nu=4/3 f$. As discussed in sec. \ref{sec:r-mode-emission}, relativistic and rotation corrections alter this simple form according to eq.~(\ref{eq:gw-frequency}). It depends on the star configuration via the compactness parameter $M/R$ but is very insensitive on the equation of state of dense matter \cite{Idrisy:2014qca}. 
For a typical neutron star the relativistic corrections are at the $10-20\%$ level \cite{Idrisy:2014qca} and rotational corrections are even smaller so that r-mode frequencies are sufficiently distinct to clearly discriminate r-modes from deformations. 

In case we have no observational information on the stars rotational frequency the discrimination of the two gravitational wave emission mechanisms is a bit harder. Since the rotation frequency can continuously vary over a large range, an observed gravitational wave frequency is generally consistent with both scenarios. Similarly the ellipticity and the r-mode amplitude are continuous parameters so that both scenarios can explain an observed gravitational wave amplitude. However, the two mechanisms differ in the qualitative generation mechanism of the gravitational waves: an ellipticity imposes a time-varying mass quadrupole, whereas r-mode oscillations impose a current quadrupole. As pointed out in \cite{Owen:2010ng}, this leads to differences in the polarization of the gravitational wave signal: the polarization angle deviates by $\pi/4$ between the two emission mechanisms and an accurate measurement of this quantity would allow us to distinguish them \cite{Owen:2010ng}.

\subsection{Direct observables}

Obviously, the most immediate insight from a gravitational wave detection is that global oscillation modes, causing the emission, are present in the star. This in turn means that strong dissipation that would damp them is absent. If the gravitational wave frequency is sufficiently low this would e.g. exclude a significant impact of Ekman layer damping at the boundary between a fluid core and a solid crust \cite{Lindblom:2000gu}, as illustrated on the left panel of fig.~\ref{fig:scenarios}. Similarly for fast spinning sources in LMXBs this could exclude the presence of a large stability window, shown on the right panel of fig.~\ref{fig:scenarios}, which would be present in various forms of matter containing ungapped quarks \cite{Madsen:1998qb}. 

The information that is further directly obtained from the gravitational wave data is the gravitational wave frequency $\nu$ and the intrinsic strain amplitude $h_0$. Actually due to the way gravitational wave searches have to be performed---namely by matched filtering, comparing the data with a detailed model for the evolution of the gravitational wave signal of a given source during the observation interval, see e.g. \cite{Wette:2008hg}---a detection will also directly yield a measurement of the rate of change of the gravitational wave frequency $\dot\nu$. Depending on the sophistication of the gravitational wave template used in the search and the quality of the data even the second time derivative $\ddot\nu$ can be obtained. For longer search intervals or rapid spindown evolution such detailed templates are even required for a detection and have already been used in previous searches \cite{Wette:2008hg}. The measured gravitational wave frequency is 
related to the rotation frequency of the star by eq.~(\ref{eq:gw-frequency}). In case the rotation frequency is not known this connection can be used to obtain an estimate for it from the gravitational wave data. Similarly, the r-mode amplitude is related to the gravitational wave amplitude via eq.~(\ref{eq:r-mode-alplitude}), yet this relation involves generally unknown source properties and can likewise only give a rough estimate.

As noted in \cite{Mytidis:2015} the combined measurement of $\nu$, $\dot\nu$ and $h_0$ can also provide a measurement for the moment of inertia of the star. In this analysis the canonical relation $\nu=4/3 f$ between the oscillation and rotation frequency was assumed. We extend this result to the realistic case of fast rotating stars with leading general relativistic corrections where the relation is slightly more complicated.
In general there can be other spindown mechanisms besides r-modes and the observed total spindown power of a source
\begin{equation}
P_{{\rm rot}}=\frac{dE_{{\rm rot}}}{dt}=I\Omega\dot{\Omega}	
\end{equation}
is larger than the r-mode component $P_R$. Using the expression for the r-mode spindown rate in terms of the gravitational wave amplitude eq.~(\ref{eq:r-mode-spindown-power}), the fraction of r-mode spindown is then
\begin{equation}
\frac{P_{R}}{P_{{\rm rot}}}=\frac{4}{15}\frac{\chi D^{2}}{I}\frac{h_{0}^{2}\Omega}{|\dot{\Omega}|}
\end{equation}
If this quantity is close to unity r-modes should dominate the spindown\footnote{Note that even when gravitational wave emission completely dominates, the ratio might not be as close to unity as expected because of errors in current distance estimates and the factor $\chi$, which in the absence of spin frequency measurements would have to be estimated by 1.}. In this case one obtains the expression for the moment of inertia
\begin{equation}
I\xrightarrow[P_{\rm rot} \approx P_{R}]{}\frac{4}{15}\frac{\chi D^{2} h_{0}^{2}\nu}{\left|\dot{\nu}\right|} \label{moment-of-inertia}
\end{equation}
This expression deviates from the analysis in \cite{Mytidis:2015} by the consideration of the factor $\chi$. If the spin frequency of the source is unknown, as is the case for newly-formed compact stars discussed in \cite{Mytidis:2015} this introduces an uncertainty of several tens of percent. In combination with the additional uncertainty in current  distance measurements this could prevent reliable results. However, in case the factor $\chi$ is measured by observation of the spin frequency, eq.~(\ref{moment-of-inertia}) does for sources with precise distance measurement indeed provide a very useful estimate on the moment of inertia. As is well known and has been pointed out in \cite{Mytidis:2015}, the moment of inertia can similarly to the radius of a source efficiently distinguish between different equations of state of dense matter which makes eq.~(\ref{moment-of-inertia}) a promising tool for compact star physics.

Finally, as known from general spindown theory \cite{1983bhwd.book.....S}, a measurement of $\ddot \nu$ could yield information on the spindown mechanism encoded in the braking index
\begin{equation}
n=\frac{f \ddot f}{\dot f^2}=\frac{\nu \ddot\nu}{\dot\nu^2} \label{eq:braking-index}
\end{equation}
Whereas the canonical r-mode braking index, obtained for a constant saturation amplitude, is 7, the fact that the saturation amplitude is for realistic saturation mechanisms a function of both temperature and frequency yields an effective spin-down law \cite{Alford:2012yn}
\begin{equation}
	 \label{eq:effective-spindown}
\end{equation}
with a braking index
\begin{equation}
n_{{\rm rm}} \equiv \frac{7+2\gamma+2\beta/\theta}{1-2\beta/\theta}
\end{equation}
that depends on the r-mode saturation mechanism and can be used to distinguish the various proposed mechanisms which can cover a large range $1\lesssim n_{\rm rm} \lesssim 10$. As known already from electromagnetic pulsar timing measurements, the second time derivative is strongly affected by disturbances of the spindown, like glitches, and therefore the gravitational wave signal might not give a reliable value for the breaking index. Moreover, in general other spindown sources should be present that contribute to the spindown and might even dominate it. Whereas for certain younger sources, like e.g. the Crab pulsar that spin too slow for r-modes to be present, reliable values are obtained from electromagnetic timing data, the situation is particularly problematic for old millisecond pulsars. Their obtained braking indices from electromagnetic observations vary in sign and range in the millions since they are strongly affected by timing irregularities \cite{Hobbs:2009vh}, that could e.g. stem from the accretion of interstellar gas, and cannot give us direct information about the spindown behavior. 

\section{Derived information from multi-messenger observations}
\label{sec:multi-messenger}

If continuous gravitational wave emission is detected it is very likely that it is associated with a known optical source. As discussed in sec. \ref{sec:detectability}, most searches are directed searches in which case the source position significantly helps in the data analysis and increases the detectability. Generally we have additional information from electromagnetic observations that can give us further insight into the source properties when combined with the information obtained directly from the gravitational wave signal.

The most basic derived information can be obtained when the frequency of the source is measured by electromagnetic pulsar observations. It is the relation of the measured gravitational wave frequency eq.~(\ref{eq:gw-frequency}) to the rotational wave frequency $\chi$. As discussed before, it deviates from the canonical Newtonian slow-rotation result result $\chi=1$ and numerical general relativistic hydrodynamic simulations with realistic equations of state show that although it depends strongly on the compactness parameter it is surprisingly insensitive to the equation of state, with deviations at the 1-2\% level. In \cite{Idrisy:2014qca} it was found that a quadratic fit\footnote{Previous post-Newtonian results \cite{Lockitch:2000aa} having a linear dependence on $M/R$ are therefore insufficient and higher order post-Newtonian corrections would be required.} gives a very precise approximation to the data of the 14 considered realistic equations of state, with the result
\begin{equation}
\chi \approx 1.029-0.059\frac{M}{R}+1.688\left(\frac{M}{R}\right)^{2}
\end{equation}
This expression holds in the slow-rotation limit and previous studies showed that rotation corrections only change the frequency connection by a few percent. Solving this for the compactness parameter yields the estimate
\begin{equation}
\frac{M}{R}\approx 0.017+0.667 \sqrt{ \frac{\nu}{f}-1.372} \label{eq:compactness-parameter}
\end{equation}
which should hold on the 10-20\% level and is therefore competitive with, if not better than, estimates from electromagnetic observations.  
Although at this point still uncertain, such relations are very interesting since both the rotation frequency and the gravitational wave frequency could be obtained with significant accuracy. 
Whereas mass measurements, that are obtained from timing observations of binary parameters in mass point approximation \cite{Miller:2015PhR}, can be very precise, radius measurements, probing the finite-size nature of a compact star and generally being obtained from noisy thermal X-ray data, are far more uncertain and only available for a few selected sources in LMXBs \cite{Guillot:2013wu}. Therefore, if relations analog to eq.~(\ref{eq:compactness-parameter}) could be derived in a controlled way at higher order \cite{Idrisy:2014qca} in a post-Newtonian \cite{Lockitch:2000aa} and slow rotation expansion \cite{Lindblom:1999yk} this would for sources with precise mass measurements provide compact star radii of unprecedented precision. In contrast to electromagnetic messengers, eq.~(\ref{eq:compactness-parameter}), relying on the the gravitational wave signal, is in particular not affected by uncertainties in distance measurements, interstellar absorption or red-shift effects. 
For very massive sources \cite{Demorest:2010bx,Antoniadis:2013pzd} high precision mass-radius pairs would strongly constrain the equation of state of dense matter and would allow to rule out most of the presently compatible models \cite{Lattimer:2006xb}. Due to the weak sensitivity of the compact star radius over a large range of presently observed, intermediate masses \cite{Lattimer:2000nx}, the results for the accessible sources in binaries would therefore strongly constrain "the typical radius of compact stars" that could then be used as an estimate to constrain the properties of other---in particular also isolated---sources. Whereas the heaviest known compact star J0348+0432 \cite{Antoniadis:2013pzd} with a period of  $f\approx 26$ Hz spins unfortunately too slow for r-modes to be present, the first confirmed heavy compact star J1614-2230 \cite{Demorest:2010bx} spins with a high enough frequency of $f\approx 317$ Hz and it presents therefore a prime target. Vice versa the compactness parameter could also provide strongly improved mass estimates for sources with radius estimates from thermal X-ray measurements. Observations of LMXBs yield generally large uncertainty areas in mass-radius space \cite{Ozel:2010fw,Steiner:2010fz} and a precise compactness parameter would favorably constrain them. Finally thermal X-ray measurements could allow a mass estimate even for nearby non-accreting \cite{Durant:2011je} or even isolated sources.

Which other information can be gained from a multi-messenger analysis depends on the additional observables that are available and therefore on the particular class of sources, discussed in section~\ref{sec:detectability}. In general, all discussed cases correspond to directed searches and additional distinct observables are obtained from electromagnetic signals that allow us to further constrain the properties of particular gravitational wave sources.

\subsection{Newly-formed compact stars}
For newly-formed sources the gravitational signal from r-modes should initially be subleading and only become dominant once the prompt gravitational wave signal from the core bounce and diffusion in the hot compact core decays \cite{Yakunin:2015wra}. Moreover, we do not have any electromagnetic spin information in this case. Yet, in addition to the gravitational wave signal we should have additional neutrino and electromagnetic information from the supernovae explosion in case it is sufficiently near. Whereas the gravitational wave signal propagates outwards immediately, the neutrino and the electromagnetic signals can be delayed. These allow us to retroactively search for the gravitational wave signal in the recorded data. Whereas there are prompt neutrinos from the supernova, neutrinos from the compact core can leave the star only gradually as the temperature drops and it becomes transparent to neutrinos, which should happen over a timescale of about a minute. The quantitative measurement of this delay would teach us about the neutrino production and transport mechanism within the star and its cooling evolution. The energies of the measured neutrinos would in addition also give an estimate of the initial temperature. Whereas the statistics was insufficient in the only case SN1987A where neutrinos were observed so far, for a galactic supernova the neutrinos should provide a good picture of the initial cooling evolution that could be correlated with the gravitational wave data to time the evolution. The electromagnetic signal is even further delayed which can provide information on the supernova dynamics. In \cite{Mytidis:2015} it has been suggested that further information about compact stars can be obtained from the initial cooling and spindown evolution. However, due to the delay and gradual start of the evolution and our ignorance on the detailed initial conditions it seems questionable if such a refined analysis is feasible.

\subsection{Young sources}
For young sources in their initial spindown phase the rotation frequency is unknown. Eq.~(\ref{eq:compactness-parameter}) therefore cannot be applied in this case and the measurement of the gravitational wave frequency eq.~(\ref{eq:gw-frequency}) yields then only an estimate on the rotation frequency of the source which is accurate within a few tens of percent. 
However, for young sources an additional observable is known from electromagnetic observations, namely the {\em age} $t$ of the source. When r-modes dominate the spindown, as necessary to explain the low frequencies of young pulsars \cite{Alford:2012yn}, the spindown is initially very fast and the observed sources spin with frequencies sufficiently below their initial frequency. In this case the spindown is independent of the initial conditions and the solution of the effective spindown equation eq.~(\ref{eq:effective-spindown}) yields then an expression for the factor $\cal C$ describing the saturation mechanism
\begin{equation}
{\cal C}=\frac{6t\left|\dot{\nu}\right|}{\nu}
\end{equation}
This expression, involving the age of the star, provides in contrast to eq.~(\ref{eq:braking-index}) a partially averaged braking index $\bar n_{\rm rm}$ including information on the entire evolution of the source\footnote{Only "partially averaged" since nevertheless the current spindown rate and frequency enter.}
\begin{equation}
\bar{n}_{\rm rm}=\frac{6}{{\cal C}}+1=\frac{\nu}{t\left|\dot{\nu}\right|}+1 \label{eq:young-braking-index}
\end{equation}
and should therefore be less dependent on temporary disturbances of the spindown. 
For recent supernovae the age is very precisely known and even for older sources, like Cas A the age is known on the percent level. The factor $\cal C$ is sufficiently distinct for known mechanisms \cite{Alford:2012yn} so that a determination of the spindown mechanism should be viable with sufficiently precise data. Actually, it is promising that eq.~(\ref{eq:young-braking-index}) can indeed provide a reliable value, since timing irregularities are far less pronounced in the spindown rate, in particular when averaged over sufficiently large time intervals, and furthermore the youngest known pulsars show timing with reliable braking indices even when using the standard expression eq.~(\ref{eq:braking-index}) \cite{Hobbs:2009vh}. For young sources the spindown mechanism might therefore be constrained independently by both expressions eqs.~(\ref{eq:braking-index}) and (\ref{eq:young-braking-index}), where the former gives the current value and the more precise latter expression a partially averaged value over the stars life. This could therefore teach us about a potential evolution of the braking index, e.g. due to the changing contributions of magnetic and gravitational wave braking during the life of the source.

\subsection{Sources in LMXBs}

For sources in LMXBs the rotation frequency is known, but the spindown rate and the age are generally unknown. By now there are first sources for which the spindown in quiescence has been observed \cite{Patruno:2010qz,Mahmoodifar:2013quw}. Unfortunately due to the complicated evolution of these sources this quantity cannot directly be connected to their age. However, for most sources in LMXBs there are X-ray measurements, i.e. we know the {\em temperature} or even the thermal luminosity from spectral fitting. The thermal data sets strict constraints on the contribution of r-modes to the spindown of those sources where quiescence spindown rates are known, since r-modes that cause a quick spindown eq.~(\ref{eq:rotational-loss}) would according to eq.~(\ref{eq:instability-power}) also strongly heat these sources. The observed temperatures show that r-modes account for less than about 1\% of the spindown in the presently considered sources \cite{Mahmoodifar:2013quw} and therefore eq~(\ref{moment-of-inertia}) unfortunately cannot be used to obtain an estimate for the moment of inertia of these sources.

However, r-modes could be far more important for the thermal state of the system. The thermal steady state in case of LMXBs is determined by $L_\gamma+L_\nu=P_{G}+P_{{\rm acc}}$ where $L_\nu$ is the neutrino luminosity, $L_\gamma$ the photon luminosity, $P_G$ is the r-mode heating eq.~(\ref{eq:-r-mode-heating}) and $P_{\rm acc}$ is the heating power due to accretion and nuclear reactions in the crust. Whereas the photon luminosity is observed in many LMXBs, the neutrino luminosity is unknown. The accretion power could be estimated within an accretion model, but we do not include such additional information here.
With the measured rotation frequency, the heating power due to r-modes eq.~(\ref{eq:-r-mode-heating}) can be expressed in terms of observable quantities
\begin{equation}
P_{G}=\frac{2\pi^{2}}{5} D^{2}\left(2f-\nu\right)\nu h_{0}^{2} \label{eq:obs-r-mode-heating}
\end{equation}
where only the distance is not directly determined by the gravitational wave signal. The photon luminosity can be expressed as
\begin{equation}
L_{\gamma}=L_{\gamma}^{\infty}\left(1-\frac{2GM}{R}\right)^{-1}=4\pi D^{2}J_{\gamma}\left(1-\frac{2GM}{R}\right)^{-1}
\end{equation}
where $L_{\gamma}^{\infty}$ is the redshifted luminosity and $J_\gamma$ is the observed photon flux in the detector. The ratio of the two is
\begin{equation}
\frac{P_{G}}{L_{\gamma}}=\frac{\pi}{10}\frac{\left(2f-\nu\right)\nu h_{0}^{2}}{J_{\gamma}}\left(1-2G\frac{M}{R}\right)
\end{equation}
The interesting aspect about this expression is that the distance drops out and all involved quantities can be precisely determined, assuming that a precise relation for the compactness parameter in terms of the frequencies analogous to eq.~(\ref{eq:compactness-parameter}) will be derived in the future. If this ratio is evaluated for a given source it can be quantitatively checked if both r-modes dominate the heating and photon emission dominates the cooling, since only in this case it should be close to one. For colder sources this is indeed a likely scenario since accretion heating must be small and photon emission would dominate over neutrino emission as long as very fast neutrino cooling processes are absent. If this identification strategy worked and the ratio for an observed source turns out to be one, the photon luminosity is determined by eq.~(\ref{eq:obs-r-mode-heating}) in terms of gravitational wave measurements. Therefore it gives an important additional constraint that can be used to improve the spectral fit, where it is required to separate the thermal X-ray component from non-thermal power-law components. This additional information could thereby improve current spectral X-ray analyses, which often have to make generic assumptions, e.g. on the radius of the source.

When the X-ray fit is e.g. well described by a black-body spectrum with luminosity
\begin{equation}
L_{\gamma}=\frac{\pi^{3}}{15}R^{2}T_{s}^{4}=\frac{\pi^{3}}{15}R^{2}\left(1-\frac{2GM}{R}\right)^{2}T_{\infty}^{4} \label{eq:black-body-luminosity}
\end{equation}
with the observed red-shifted temperature $T_\infty$ one can obtain an expression for the radius of the source
\begin{equation}
R \xrightarrow[L_{\gamma}\gg L_{\nu},P_G \gg P_{\rm acc}]{} \sqrt{\frac{6}{\pi}\left(2f-\nu\right)\nu}\frac{Dh_{0}}{T_{\infty}^{2}}\left(1-2G\frac{M}{R}\right)^{-1} \label{eq:radius}
\end{equation}
This expression can be compared to the standard radius determination obtained from eq.~(\ref{eq:black-body-luminosity}) when the thermal luminosity is extracted from the observed X-ray flux \cite{Ozel:2010fw,Steiner:2010fz}. Whereas the electromagnetic signal is strongly affected by interstellar absorption, expressed by the poorly-constrained hydrogen column density parameter, the gravitational wave signal is unaffected by absorption due to interstellar gas. Our ignorance on the precise value of the hydrogen column density can induce an uncertainty of up to a factor 2 on the radius. Therefore eq.~(\ref{eq:radius}) could eliminate a significant element of uncertainty. With the help of the compactness parameter eq.~(\ref{eq:compactness-parameter}) the radius eq.~(\ref{eq:radius}) would then also provide masses for isolated young or nearby X-ray sources.

In case the radius is already independently determined by a precise mass measurement and the compactness parameter, the relation can alternatively be resolved for the least constrained quantity, which is the distance to the source. This way it would provide a direct distance measurement
\begin{equation}
D \xrightarrow[L_{\gamma}\gg L_{\nu},P_G \gg P_{\rm acc}]{} \sqrt{\frac{\pi}{6\left(2f-\nu\right)\nu}}\frac{RT_{\infty}^{2}}{h_{0}}\left(1-\frac{2GM}{R}\right) \label{eq:distance}
\end{equation}
which would be welcome taking into account that many distance measurements have significant uncertainties. If the accuracy of the constrained thermal fitting would be improved, eq.~(\ref{eq:distance}) could be reasonably accurate to compete with other methods like cepheids. LMXBs are particularly frequent in globular clusters. Therefore, this distance estimate would provide a way to gauge present distance estimates for these sources which are an important link in the cosmic distance ladder. In particular, eq.~(\ref{eq:distance}) could yield direct distance measurements for sources that are more than 10 kpc away without relying on "lower ladder spokes". With future very precise interferometers this might even be expandable to more distant sources  and pulsars could therefore become precise "standard sirens" for astronomy.

\subsection{Millisecond pulsars}

For millisecond pulsars there is a wealth of very precise {\em timing data}. As already discussed in sec.~\ref{sec:detectability}  this strongly simplifies gravitational wave searches and should due to the strongly constrained data analysis also provide more precise gravitational wave observables, like the amplitude and the frequency. Given a higher order analysis within a post-Newtonian and slow-rotation expansion this would yield a ultra-precise measurement of the compactness parameter analog to eq.~(\ref{eq:compactness-parameter}). As discussed in sec. \ref{sec:detectability}, in order for r-mode gravitational wave emission from isolated millisecond pulsars to be detectable with second or third generation detectors would require that r-modes dominate the spindown of these sources. And in contrast to LMXBs, there is no data for millisecond pulsars that would exclude this \cite{Alford:2013pma}. In particular, as discussed in the last section, the braking indices are affected by timing irregularities \cite{Hobbs:2009vh} and therefore cannot give us a any clue about the spindown mechanism. The spindown power due to r-modes expressed in observable quantities is from eq.~(\ref{eq:r-mode-spindown-power})
$P_{R}=4\pi^{2}/5 D^{2} f\nu h_{0}^{2} \label{eq:obsr-mode-spindown-power}$,
which can be used to check if r-modes indeed dominate the spindown. If so, we can derive further properties about the source from the measured gravitational wave amplitude. Inserting the spindown rate from electromagnetic timing observations in eq.~(\ref{moment-of-inertia}) provides a measurement of the moment of inertia
\begin{equation}
I \xrightarrow[P_{\rm rot} \approx P_{R}]{}\frac{D^{2}\nu h_{0}^{2}}{5 \left|\dot{f}\right|} \label{obs-moment-of-inertia}
\end{equation}
Correspondingly with eqs.~(\ref{eq:compactness-parameter}) and (\ref{obs-moment-of-inertia}), two of the three important bulk parameters of a source $M$, $R$ and $I$ are determined by gravitational wave measurements. For realistic neutron star equations of state these quantities are correlated \cite{Lattimer:2000nx} which should allow to estimate the possible mass and radius ranges. Moreover, if an estimate for the narrow range of compact star radii at presently-observed, intermediate masses is obtained from improved future X-ray observations or independent gravitational wave observations in LMXBs, as discussed at the beginning of this section, this would then also allow to give a direct estimate for the mass of an isolated millisecond pulsar using the measured compactness parameter.

Similarly the observed gravitational wave amplitude can also be used to give an estimate on the r-mode amplitude via eq.~(\ref{eq:r-mode-alplitude}), which yields $\alpha\!=\!\sqrt{3^{2}5^{3}/(2^{13}\pi^{7}G)}$ $\times Dh_{0}/(\tilde{J}MR^{3}\nu^{2}f)$.
Even if the mass and the radius can be obtained from multi-messenger observations, this leaves the dependence on the unknown radial profile parameter $\tilde J$ eq.~(\ref{eq:J-tilde}) which is again correlated with the other bulk parameters. It has been shown \cite{Alford:2012yn} that this parameter is generally bounded within a factor 2 , and takes values around $\tilde J \approx 0.02$, so that even without further input the above expression yields a rough estimate on the r-mode amplitude. 

For millisecond pulsars, where generally no thermal estimates are available, the thermal heating power eq.~(\ref{eq:obs-r-mode-heating}) can finally be used to obtain a surface temperature estimate. In isolated millisecond pulsars where there is no accretion from a companion there should be no other strong heating source and therefore an r-mode producing a detectable gravitational signal should dominate the heating. Correspondingly, these sources are expected to be rather cold so that photon cooling from the surface should dominate. Under this assumption one can, analogously to the discussion in the last subsection, obtain an estimate for the surface temperature
\begin{equation}
T_{s}\xrightarrow[P_G \approx L_\gamma]{}\left(\frac{6}{\pi}\frac{D^{2}\left(2f-\nu\right)\nu h_{0}^{2}}{R^{2}}\right)^{\frac{1}{4}}
\end{equation}
This estimate is similar to the r-mode temperature \cite{Alford:2013pma} obtained from the electromagnetic timing data. However, it does not require that r-modes dominate the spindown, but only the weaker requirement\footnote{Note that potential other spindown mechanisms like magnetic dipole emission would not heat the star.} that they dominate the heating of the source.


\section{Conclusion}

We conclude that r-mode astronomy presents a unique way to study properties of compact stars ranging from bulk parameters to the composition of their opaque interior, and would provide us with precision information that would in many cases be virtually impossible to obtain otherwise. In particular the combination of gravitational data with current and future electromagnetic observations \cite{Nandra:2013shg,2013MmSAI..84..782S,2012SPIE.8443E..13G} could significantly enhance our understanding of compact stars. Direct expressions in terms of gravitational wave and electromagnetic observables are given for the compactness parameter by eq.~(\ref{eq:compactness-parameter}), the moment of inertia when r-modes dominate the spindown eq.~(\ref{obs-moment-of-inertia}), and the radii or distances of old sources for which r-modes dominate the heating and the observed photon luminosity dominates the cooling eqs.~(\ref{eq:radius}) and (\ref{eq:distance}).

R-mode astronomy obviously relies on the presence of r-modes in compact stars. Most scenarios on the damping within the star predict r-mode emission in at least part of the discussed source classes. For instance the minimal viscous damping in ordinary neutron stars is insufficient to completely damp r-modes, so that they should be present in all four classes discussed above for saturation amplitudes in the range predicted by current saturation mechanisms. Superfluidity and superconductivity could have a big impact on the dissipation in older sources, but they should be absent in young sources due to their large temperatures. Similarly, exotic forms of matter like quark or hyperonic matter that impose an r-mode stability window could completely damp r-modes in low mass X-ray binaries, yet they should nevertheless be present in old isolated millisecond pulsars.

Finally even though it would be very unfortunate if we would not detect gravitational waves from some of the above classes of sources, particularly when taking into account the significant insight that can be gained from such a detection, such a "negative result" would still be interesting. Since this would be inexplicable in standard neutron stars with minimal viscous damping, it would require an explanation in terms of very strong dissipation within the star and would therefore inform us about potential exotic forms of dense matter realized there \cite{Alford:2010fd,Haskell:2012,Alford:2013pma}.

\section*{Acknowledgements}

It is our pleasure to thank Kostas Glampedakis and Tolga G\"uver for very helpful discussions.

%
%

%
%
%
%

\end{document}